\documentclass[preprint,11pt]{JHEP3}

\JHEPspecialurl{http://jhep.sissa.it/JOURNAL/JHEP3.tar.gz}

\usepackage{epsfig,multicol,amsmath}

\usepackage{epstopdf}
\usepackage{bm}  
\usepackage{amsmath}     
\usepackage{epsfig,epsf}
\usepackage{rotating}
\usepackage{cite}

\def\beq{\begin{equation}}
\def\eeq{\end{equation}}
\def\bea{\begin{eqnarray}}
\def\eea{\end{eqnarray}}

\def\eq#1{{Eq.~(\ref{#1})}}
\def\fig#1{{Fig.~\ref{#1}}}
\newcommand{\bas}{\bar{\alpha}_S}
\newcommand{\as}{\alpha_S}

\newcommand{\Lb}{\left(}
\newcommand{\Rb}{\right)}
\newcommand{\h}{\frac{1}{2}}
\parskip=\itemsep               

\newcommand{\nn}{\nonumber}

\title{Geometric scaling behavior of the scattering amplitude  for DIS with nuclei}
\author{\Large Andrey Kormilitzin${}^{a}$\thanks{Email: andreyk1@post.tau.ac.il.},\,\,\,
Eugene\, Levin${}^{a, b}$ \thanks{Email: leving@post.tau.ac.il, eugeny.levin@usm.cl.}\,\,\,and\,\,\, Sebastian Tapia${}^b$ \thanks{Email: trockut@gmail.com}
\\
${}^a$ \, Department of Particle Physics, School of Physics and Astronomy,
Tel Aviv University, Tel Aviv, 69978, Israel\\
${}^b$\, Departamento de F\'\i sica,
Centro de Estudios Subat$\acute{o}$micos,
Universidad T$\acute{e}$cnica Federico Santa Mar\'\i a,\\ and
Centro Cient´ıfico-Tecnol$\acute{o}$gico de Valpara\'\i so,
Casilla 110-V,  Valparaiso, Chile\\
}

\abstract
{The main question, that we answer in this paper, is whether the initial condition can influence  on  the geometric scaling behavior of the amplitude for DIS at high energy. We re-write the non-linear Balitsky-Kovchegov equation in the form which is useful for treating the interaction with nuclei.
Using the simplified BFKL kernel, we find the analytical solution to this equation with the initial condition given by the McLerran-Venugopalan formula.  This solution does not show the geometric scaling behavior of the amplitude deeply in the saturation region. On the other hand, the BFKL Pomeron calculus with the initial condition  at $x_A = 1/mR_A$ given by the solution to Balitsky-Kovchegov equation, leads to the geometric scaling behavior.
The McLerran - Venugopalan formula is the natural initial condition for the Color   Glass Condensate (CGC) approach. Therefore, our result gives a possibility to check experimentally which approach: CGC or BFKL Pomeron calculus, is more adequate. }

\keywords{Color Glass Condensate, gluon saturation, BFKL Pomeron, calculus,  non-linear evolution, geometric scaling behavior
}
\dedicated{PACS: 12.38-t, 12.38.Cy,1 2.38.Lg, 13.60.Hd, 24.85.+p, 25.30.Hm}
\preprint{TAUP 2929/11 \\
{\tt }\\
\today}

\begin{document}
\section{Introduction.}
Geometric scaling behavior of the scattering amplitude is one of the most qualitative and well founded features of the high energy scattering in the framework of the CGC approach. In the saturation domain it follows from the general structure of the  non-linear equation (see Ref.\cite{BALE}) that governs the scattering amplitude in high density QCD \cite{GLR,MUQI,MV,JIMWLK,B,K}.
In the vicinity of the saturation scale the geometric scaling behavior was derived \cite{IIM} from the linear evolution \cite{DGLAP,BFKL}. The geometric scaling behavior means that the amplitude turns out to be a function of one dimensionless variable: $\tau = r^2 Q^2_s\Lb x; b \Rb$ instead of  three variables: $x$ - the fraction of the energy carried by the parton interacting with the virtual photon in DIS, $r$- the typical size of the interacting dipole ($r \propto 1/Q$ where $Q$ is the photon virtuality), and $b$ is the impact parameter of the scattering process.  Actually, the geometric scaling behavior reflects that the only dimensional parameter , that governs the scattering process in the saturation domain,  is the saturation scale $Q_s$ and $\tau$ is just the only dimensionless variable that we can construct.  Implicitly, such a behavior assumes that the amplitude in the saturation region does not depend on the initial condition for the non-linear equation. As we have mentioned  this idea looks consistent with the geometric scaling behavior of the scattering amplitude in the perturbative QCD region near to the saturation scale \cite{IIM}.

However, it has been noticed in Ref. \cite{LTAA} that the geometric scaling behavior of the scattering amplitude cannot be correct  for the DIS with nuclei if the McLerran-Venugopalan formula \cite{MV}  is used as the initial condition for this process. This initial condition itself shows the saturation behavior on the scale which is the saturation scale at low energy ($Q_s\Lb x=x_0;b\Rb$). Generally , this behavior could be different from the geometric scaling one as it is shown in Ref. \cite{LTAA}.

In this paper we revisit this problem and we try to answer the following questions:(i) could the initial condition affect the behavior of the scattering amplitude at $\tau \,\gg\,1$; (ii) can we trust the McLerran-Venugopalan formula  deeply in the saturation region; and  (iii) what initial condition we need to use to reproduce the geometric scaling behavior.

We believe that all these questions are needed to be answered not only due to pure theoretical interest but also because the geometric scaling behavior is seen experimentally both for hadron and nucleus scattering (see Refs.\cite{GSEXP}).
\section{General solution to the simplified non-linear equation.}

The nonlinear  Balitsky-Kovchegov equation for the scattering amplitude of the dipole with size $r$ has the following form\cite{B,K}:
\bea
\frac{\partial N\Lb r,Y;b \Rb}{\partial\,Y}\,
\,\,&
=&\,\,\frac{C_F\,\as}{2\,\pi^2}\,\,\int\,d^2 r'\,K\Lb r; r'\Rb \times
\Big\{2  N\Lb r',Y;\vec{b} \,-\,
\frac{1}{2}\,(\vec{r} - \vec{r}\:')\Rb\,\,\label{MEQ1}\\
 &-& \,\, N\Lb r,Y;\vec{b}\Rb\,\,- \,\,
 N\Lb r',Y;\vec{b} - \frac{1}{2}\,(\vec{r} - \vec{r}\:') \Rb\, N\Lb
 \vec{r} -
\vec{r}\:',Y;b - \frac{1}{2} \vec{r}\:'\Rb \Big\} \nn
\eea
where $Y = \ln(1/x)$ is the rapidity of the incoming dipole; $N$ is the imaginary part of the scattering amplitude and $b$ is the impact parameter of this scattering process. $\as $ is the QCD coupling, below we will use the following notation $\bas = \as N_c /\pi$. The BFKL kernel $K\Lb r; r'\Rb$ has the following form
\beq \label{K}
 K\Lb r; r'\Rb\,\,=\,\,\frac{r^2}{(\vec{r}\,-
\,\vec{r}\:')^2\,r'^2}\,
\eeq

\subsection{Simplified equation for scattering with nuclei}

Considering the interaction of the dipole with a nucleus we can simplify this equation. Indeed, neglecting the non-linear term in \eq{MEQ1} we have the linear BFKL equation and the amplitude for dipole-nucleus scattering can be written as

\beq \label{MEQ2}
N^{BFKL}_A = \int d^2  b'\,
N^{BFKL}_N\Lb r,\,Y;\,\vec{b} \,-\, \vec{b}' \Rb \,T_A(b') =T_A(b)\,\int d^2 b"\,N_N(r,Y;\,b")\,=\,N_N\Lb r,Y;t=0\Rb T_A(b)
\eeq
where $ \vec{b} \,-\, \vec{b}' \,\equiv\,\vec{b}"$ and 
$T_A(b)$ is  the number of the nucleons at given impact parameter $b$ in the nucleus. It can be calculated as the following integral
\beq \label{MEQ3}
T_A(b)\,\,=\,\,\int^{+ \infty}_{-\infty}  d z\, \rho\Lb b, z\Rb
\eeq
where $\rho$ is the density of the nucleons in a nucleus and $z$ is the longitudinal coordinate of the nucleon.
This number depends on $b$ but it is of the order of $A^{1/3}$.

In \eq{MEQ2} we assumed that $|\vec{b} - \vec{b}'| \leq R_N \,\ll\,b' \approx R_A$. One can recognize that this is
typical Glauber-type assumption which holds for interaction with  nuclei if the energy is not so high that the radius dipole-nucleon interaction can be considered as much smaller than the nucleus radius. $N_N\Lb r,Y;t=0\Rb$ is the imaginary part of dipole-nucleon amplitude at momentum transfer $t=0$, which proportional to dipole-nucleon total cross section.

 Let us solve \eq{MEQ1} using \eq{MEQ2} as the first iteration. One can see that the second iteration leads to the following expression
\bea \label{MEQ4}
\frac{\partial N^{(2)}_A\Lb r,Y;t=0 \Rb}{\partial\,Y}\,
\,\,&
=&\,\,\frac{\bas}{2\,\pi}\,\int\,\frac{d^2 r'\,r^2}{(\vec{r}\,-
\,\vec{r}\:')^2\,r'^2}\, \times
\Big\{2\,T_A(b) \,  N^{BFKL}_N\Lb r',Y;t=0)\Rb\,\\
 &-& \,\,T_A(b)\, N^{BFKL}_N\Lb r,Y;t=0\Rb\,\,- \,\,T^2_A(b)\,
 N^{BFKL}_N\Lb r',Y;t=0) \Rb\, N^{BFKL}_N\Lb
 \vec{r} -
\vec{r}\:',Y; t=0\Rb \Big\} \nn
\eea

Introducing $N^{eff}_N = N_A/T(b)$ one can see that \eq{MEQ4}  can be viewed  as the first iteration of the equation
\bea \label{MEQ5}
\frac{\partial N^{eff}_N\Lb r,Y;t=0\Rb}{\partial\,Y}\,
\,\,&
=&\,\,\frac{\bas}{2\,\pi}\,\int\,\frac{d^2 r'\,r^2}{(\vec{r}\,-
\,\vec{r}\:')^2\,r'^2}\, \times
\Big\{2\, N^{eff}_N\Lb r',Y;t=0)\Rb\,\\
 &-& \,\, N^{eff}_N\Lb r,Y;t=0\Rb\,\,- \,\,T_A(b)\,
 N^{eff}_N\Lb r',Y;t=0) \Rb\, N^{eff}_N\Lb
 \vec{r} -
\vec{r}\:',Y; t=0\Rb \Big\} \nn
\eea

The non-linear equation in the form of \eq{MEQ5} was firstly proposed in Refs. \cite{GLR,MUQI} and it is valid for energies
at which the radius of dipole - nucleon interaction is much smaller than the nucleus radius.  The energy dependence of radius of dipole - nucleon interaction unfortunately cannot be solved in the framework of the Balitsky-Kovchegov non-linear equation of \eq{MEQ1} (see Ref.\cite{KOWI})  since it demands some additional input from the non-perturbative QCD.  The high energy phenomenology based on the soft Pomeron approach leads to
\beq \label{MEQ6}
R^2\Lb \mbox{dipole - nucleon} \Rb \,=\,\,R^2_0 + \alpha'_P(0)  \,Y  \,\,\ll\,\,R^2_A
\eeq
in wide range of rapidity $Y$ including the LHC energies for $R^2_0 = 10 \,GeV^{-2}$ and $\alpha'_P(0) \leq 0.2 \,GeV^{-2}$.

\subsection{Simplified  BFKL kernel}
The BFKL kernel of \eq{K} is rather complicated and the analytical solution with this kernel has not been found. In Ref.\cite{LT} it was suggested to simplify the kernel by taking into account only log contributions. From formal point of view this simplification means that we consider only leading twist contribution to the BFKL kernel. Notice that the kernel of \eq{K} includes all twists contributions. We are dealing with two kinds of logs in \eq{MEQ1}, which corresponds two different kinematic regions: $\tau\,<\,1$ and $\tau\,>\,1$.

\begin{boldmath}
\subsubsection{$\tau\,<\,  1$}
\end{boldmath}

 In this kinematic region we can simplify  $K\Lb r;r'\Rb$ in \eq{MEQ1}  in the following way\cite{LT}, since $r'
 \gg r$ and $|\vec{r} - \vec{r}'| > r$
 \beq \label{K1}
 \int d^2 r' \,K\Lb r, r'\Rb\,\,\rightarrow\,\pi\, r^2\,\int^{\frac{1}{\Lambda^2_{QCD}}}_{r^2} \frac{ d r'^2}{r'^4}
\eeq
Introducing $n^{eff}_N \Lb r,Y;t=0\Rb= N^{eff}_N\Lb r,Y;t=0\Rb/r^2$ we obtain
 \beq \label{MEQ7}
\frac{\partial^2 n^{eff}_N\Lb r,Y;t=0\Rb}{\partial Y\,\partial \ln\Lb 1/(r^2 \Lambda^2_{QCD})\Rb}\,\,\,=\,\,\frac{\bas}{2}\,\Big( 2 n^{eff}_N\Lb r,Y;t=0\Rb\,\,- \,\,r^2\Lambda^2_{QCD} \,T_A(b)\,\Lb n^{eff}_N\Rb^2\Lb r,Y;t=0\Rb\Big)
\eeq
One can see that  the simplified kernel of \eq{K1} sums $\Big(\bas \ln\Lb r^2\,\Lambda^2_{QCD}\Rb\Big)^n$.
\begin{boldmath}
\subsubsection{$\tau\,>\,1$}
\end{boldmath}

The main contribution in this kinematic region originates from the decay of the large size dipole into one small size dipole  and one large size dipole.  However, the size of the small dipole is still larger than $1/Q_s$. This observation can be translated in the following form of the kernel
\beq \label{K2}
 \int d^2 r' \,K\Lb r, r'\Rb\,\,\rightarrow\,\pi\, \int^{r^2}_{1/Q^2_s(Y,b)} \frac{ d r'^2}{r'^2}\,\,+\,\,
\pi\, \int^{r^2}_{1/Q^2_s(Y,b)} \frac{ d |\vec{r} - \vec{r}'|^2}{|\vec{r} - \vec{r}'|^2}
\eeq
One can see that this kernel leads to the $\ln\Lb r^2Q^2_s\Rb$-contribution. Introducing a new function $\tilde{N}^{eff}_N \Lb r,Y;t=0\Rb\,\,=\,\,\int^{ r^2} d r^2\,N^{eff}_N\Lb r,Y;t=0\Rb/r^2$ one obtain the following equation
\beq \label{MEQ8}
\frac{\partial^2 \tilde{N}^{eff}_N\Lb r,Y;t=0\Rb}{ \partial Y\,\partial \ln r^2}\,\,=\,\, \bas \,\left\{ \Lb 1 \,\,-\,T_A\Lb b\Rb\,\frac{\partial \tilde{N}^{eff}_N\Lb r,Y;t=0 \Rb}{\partial  \ln r^2}\Rb \, \tilde{N}^{eff}_N\Lb r,Y;t =0\Rb\right\}
\eeq

The Mellin transform of the full BFKL kernel of \eq{K} has the form
\beq \label{KML}
\chi\Lb \gamma\Rb \,\,=\,\,\int \frac{d \xi}{2 \pi i}\,e^{- \gamma \xi}\,K\Lb r; r'\Rb\,\,=\,\,
2 \psi(1) \,- \,\psi(\gamma) \,-\,\psi(1 - \gamma)
\eeq
where $\xi \,=\,\ln(r^2/r'^2)$ and $\psi(z) = d\ln \Gamma(z)/d z$ with $\Gamma(z)$ equal to Euler gamma function.
The simplified kernel replaces \eq{KML} by the following expression
\bea \label{KSM}
\chi\Lb \gamma\Rb\,\,=\,\,\left\{\begin{array}{l}\,\,\,\frac{1}{\gamma}\,\,\,\,\,\mbox{for}\,\,\,\tau\geq \,1\,;\\ \\
\,\,\,\frac{1}{1 \,-\,\gamma}\,\,\,\,\,\mbox{for}\,\,\,\tau\,\leq\,1\,; \end{array}
\right.
\eea
One can see that the advantage of the simplified kernel of \eq{KSM} is that it provides a matching with the DGLAP evolution equation\cite{DGLAP} in Double Log Approximation (DLA) for $\tau < 1$.
We will show below that this kernel leads to the geometric scaling behavior of the scattering amplitude. The other attempt\cite{MUPE} to
use a simplified kernel is related to the BFKL kernel in the diffusion approximation,namely,
\beq \label{KDA}
\chi(\gamma)\,\,=\,\,\omega_0\,\,+ D\,\Lb \gamma - \frac{1}{2}\Rb^2\,\,+\,\,{\cal O}\Lb \Lb\gamma - \frac{1}{2}\Rb^3\Rb
\eeq
with
\beq \label{KDA1}
\omega_0\,\,=\,\,4\,\bas\ln 2\,;\,\,\,\,\,\,\,\,\,D\,\,=\,\, 14 \bas \zeta(3)
\eeq
In this approach we loose any matching with the GLAP evolution. Both simplified kernels reproduce the geometric scaling behavior giving the illustrations to the general conclusions of Ref.\cite{BALE}.
\subsection{Traveling wave solution and the geometric scaling behavior of the scattering amplitude.}
It is well known (see Refs.\cite{GLR,BALE,MUPE,MUTR})  that the equation for the saturation scale does not depend on the particular form of the non-linear term in \eq{MEQ1} and it has the form
\beq \label{TR1}
\ln\Lb Q^2_s(Y)/Q^2(Y=Y_0)\Rb\,\,=\,\,\frac{\chi\Lb \gamma_{cr}\Rb}{1 - \gamma_{cr}}\,\Big( Y \,\,-\,\,Y_0\Big)
\eeq
with the critical anomalous dimension $\gamma_{cr}$ given by
\beq \label{GACREQ}
- \frac{\partial \omega(\gamma_{cr})}{ \partial \gamma_{cr}}\,\,=\,\,\frac{\omega(\gamma_{cr})}{ 1 - \gamma_{cr}}
\eeq
Inserting \eq{KML} in \eq{GACREQ} one obtains
\beq \label{GACR}
\gamma_{cr}\,\,=\,\,\frac{1}{2}\;\;\;\;\mbox{and}\;\;\;\;\ln\Lb Q^2_s(Y)/Q^2(Y=Y_0)\Rb\,\,=\,\,4 \bas \Lb Y - Y_0\Rb
\eeq

In the vicinity of the saturation scale $ \tau \to 1$ the behavior of the dipole amplitude has the form \cite{MUTR,IIM}
\beq \label{VICQS}
N\Lb Y; r\Rb\,\,\propto \Lb r^2 Q^2_s\Rb^{ 1 - \gamma_{cr}}
\eeq
We illustrate this behavior  approaching to the saturation scale from the perturbative
QCD region ($\tau \ll 1$).  In this region we can use \eq{MEQ7} neglecting the non-linear term.  This equation has a simple DLA solution
\beq \label{TR2}
n^{eff}_N\,\,=\,\,\exp\Big( 2 \sqrt{\bas\,\Lb Y - Y_0 \Rb\,\ln\Lb r^2\,Q^2_s\Lb Y = Y_0 \Rb\Rb}\Big)
\eeq
 which leads to the following expression for dipole -nucleus amplitude (see \eq{MEQ2})
\bea
N_A\,\,&=&\,\,T_A\Lb b \Rb\, \exp\Big( 2 \sqrt{\bas \Lb Y - Y_0\Rb\,\ln\Lb 1/\Lb r^2\,Q^2_s\Lb Y = Y_0 \Rb\Rb\Rb} \,+\,\ln\Lb r^2 \,\,Q^2_s\Lb Y = Y_0\Rb\Rb\Big)\,\nn\\
&\xrightarrow{\tau \to 1}&\,T_A\Lb b \Rb\,\Lb r^2 Q^2_s\Rb^{ 1/2}\,\exp\Big(
- \frac{\ln^2 \tau}{ 8 \ln\Lb Q^2_s\Lb Y\Rb/Q^2_s\Lb Y = Y_0\Rb\Rb}\Big)
\eea

Therefore, in vicinity of the saturation scale if $\ln \tau \ll \sqrt{8  \ln\Lb Q^2_s\Lb Y\Rb/ Q^2_s\Lb Y = Y_0\Rb\Rb}$
the dipole-nucleus amplitude can be written as
\beq \label{TR3}
N_A\,\,=\,\,T_A\Lb b \Rb\,\phi_0\,\Lb r^2 Q^2_s\Rb^{ 1/2}\,\,=\,\,\phi_0 e^{\frac{1}{2} z}
\eeq
where
\beq \label{Z}
z\,\,=\,\,4\,\bas \Lb Y - Y_0\Rb \,+\,\ln\Lb r^2 Q^2_s\Lb A; Y=Y_0\Rb \Rb \,\,=\,\,\xi_s\,\,+\,\xi
\eeq
where
\beq \label{KSI}
\xi_s\,=\,\ln \Big( Q^2_s\Lb A;Y\Rb/Q^2_s\Lb A; Y = Y_0\Rb\Big)\,=\,4 \bas \Lb Y - Y_0\Rb,\;\;\;\;\mbox{and}\;\;\;\xi\,=\,\ln \Big( r^2 Q^2_s\Lb A; Y=Y_0\Rb \Big)
\eeq
The saturation scale for nucleus we defined as $ Q_s^2\Lb A,Y=Y_0\Rb\,\,=\,\,T^2_A\Lb b \Rb\,Q_s\Lb N;Y=Y_0\Rb$
where $Q_s\Lb N;Y=Y_0\Rb$ is the saturation scale for the nucleon at the initial energy. $\phi_0$ is a constant that absorbers all pre-exponential factors in the DLA solution. It is instructive to notice that $\phi_0 \propto \bas$\cite{GLR,LT}.

Inside the saturation region we  are looking for the solution of \eq{MEQ8} in the form\cite{LT,LTAA}:
\beq \label{NSAT}
\widetilde{N}^{eff}_N\,\,=\,\,T^{-1}_A(b)\int^{\xi}_{\xi_s}
 d \xi'\,\Big( 1\,-\,e^{ - \phi(\xi',Y)}\Big)
\eeq

From \eq{NSAT} one can see that we can easily to calculate the dipole-nucleus amplitude
\beq \label{NASAT}
N_A\,\,=\,\,T_A\Lb b \Rb N^{eff}_N\,\,=\,\,1\,\,-\,\,e^{- \phi\Lb \xi,Y\Rb}
\eeq
Substituting \eq{NSAT} into \eq{MEQ8} we obtain
\beq \label{NSAT1}
 \phi'_Y\,e^{ - \phi}\,\,=\,\, \bas \widetilde{N}^{eff}_N\,e^{ - \phi}
\eeq
Canceling $e^{ - \phi}$ and differentiating with respect to $\xi$ we  obtain the equation in the form:
\beq \label{EQXIY}
\frac{\partial^2 \phi}{ \partial Y\,\partial \xi}\,\,=\,\,\,\bas\,\Big( 1\,-\,e^{ - \phi\Lb Y;\xi\Rb}\Big)
\eeq
Using variable $\xi_s$ and $\xi$ we can rewrite \eq{NSAT1} in the form
\beq \label{EQXIXIS}
\frac{\partial^2 \phi}{ \partial \xi_s\,\partial \xi}\,\,=\,\,\frac{1}{4}\Big( 1\,-\,e^{ - \phi\Lb Y;\xi\Rb}\Big)
\eeq
or in the form of
\beq \label{EQZX}
\frac{\partial^2 \phi}{ \partial  z^2}\,\,-\,\,\frac{\partial^2 \phi}{ \partial  x^2}\,\,=\,\,\frac{1}{4}\Big( 1\,-\,e^{ - \phi\Lb Y;\xi\Rb}\Big)
\eeq
for $z$ defined in \eq{Z} and $x= \,\xi_s - \xi$.

\eq{EQZX} has general traveling wave solution (see Ref.\cite{MATH} formula {\bf 3.4.1})

\beq \label{GSOL}
\int^\phi_{\phi_0}\frac{d \phi'}{\sqrt{c \,+\,\frac{1}{2 ( \lambda^2 - \kappa^2)}\Big( \phi'  - 1 + e^{-\phi'}\Big)}}\,\,=\,\,  \kappa \,x +\lambda \,z
\eeq
where $c, \phi_0,\lambda$ and $\kappa$ are arbitrary constants that should be found from the initial and boundary conditions.

From the matching with the perturbative QCD region (see \eq{TR3}) we have the following initial conditions:
\beq \label{IC}
\phi\Lb t \equiv z = 0, x\Rb\,\,=\,\,\phi_0\,;\,\,\,\,\,\,\phi'_z\Lb t \equiv z = 0, x\Rb\,\,=\,\,\frac{1}{2}\,\phi_0
\eeq

These conditions allow us to find that $\kappa=0$ and  $c=0$ for $\phi_0 \,\ll\,1$.
 Therefore, solution of \eq{GSOL} leads to the geometric scaling since it depends only on one variable:   $z$.     \,It has the form\cite{LT}
\beq \label{SOL}
\sqrt{2}\int^\phi_{\phi_0}\frac{d \phi'}{\sqrt{\phi'  - 1 + e^{-\phi'}}}\,\,=\,\, z
\eeq
For arbitrary $\phi_0$ the solution has the form
\beq \label{SOLL}
\int^\phi_{\phi_0}\frac{d \phi'}{\sqrt{\frac{1}{4}\,\phi_0^2\,+\,\frac{1}{2} \Big(  \phi' \,-\,\phi_0 \,+ \,e^{-\phi'}\,-\,e^{- \phi_0}\Big)}}\,\,=\, \,z
\eeq


\FIGURE[h]{\begin{minipage}{7cm}{
\centerline{\epsfig{file=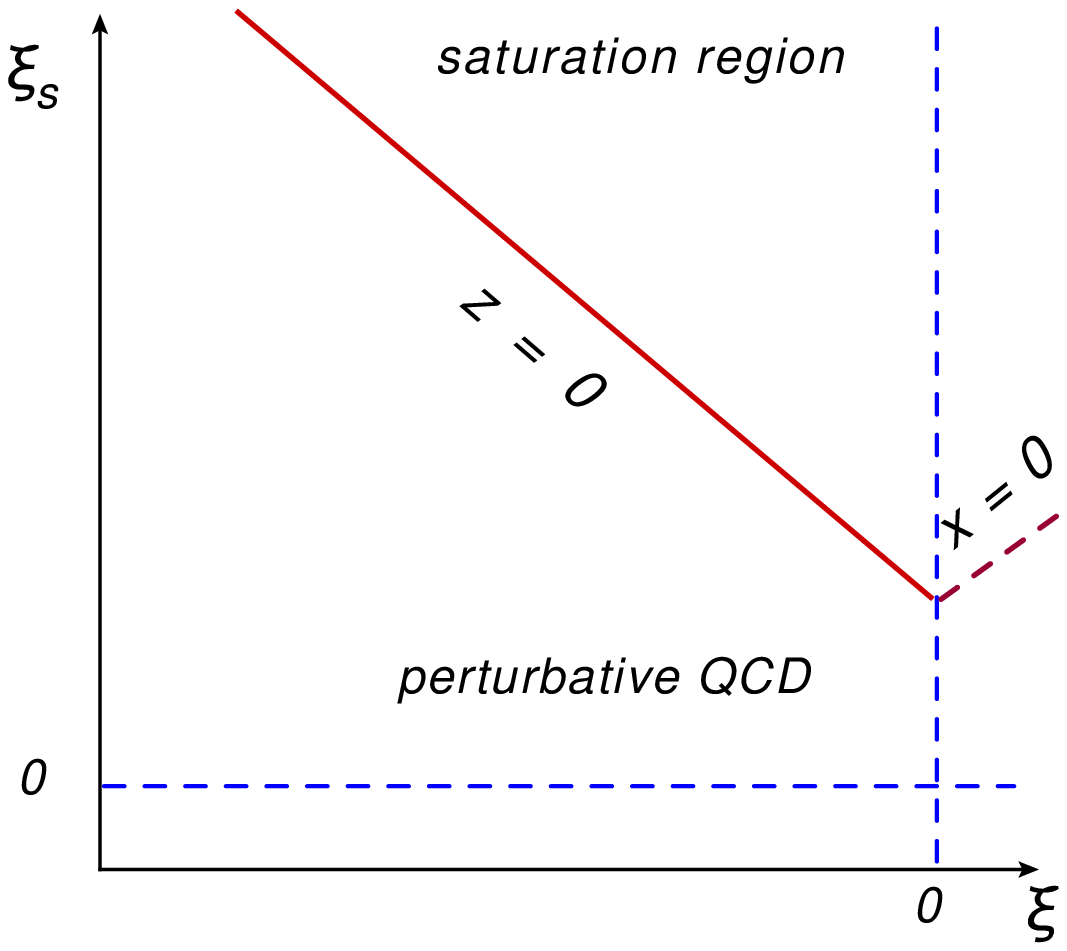,width=65mm}}}
\end{minipage}
\begin{minipage}{6cm}{
\caption{Saturation region: $z = \xi_s + \xi$ and $x = \xi_s - \xi$ for interaction with dilute target (proton)}
\label{1}}
\end{minipage}
}
Solution of \eq{SOL} we can use in the case of interaction of the dipole with rather dilute target. In this case the initial
conditions are determined by perturbative QCD and have the form of \eq{IC}.

One can see that if $\phi_0 \propto \bas \,\ll\,1$ the solution of \eq{SOL} gives
\beq \label{SOL1}
\phi\,\,=\,\,\phi_0 e^{\h z}
\eeq
while at $z \gg 1$ \eq{SOL} leads to
\beq \label{SOL2}
\phi \,\,=\,\,\frac{z^2}{8}
\eeq

~

~
\section{Dipole-nucleus amplitude: solution for one  critical line and violation of the geometric scaling behavior.}
The main ingredient of Color Glass Condensate (CGC) approach is the assumption that there exists such value of energy that we can describe dipole-nucleus amplitude using the McLerran-Venugopalan formula\cite{MV}:
\beq \label{ADA1}
N_A\Lb r^2;Y; b \Rb\,\,\,=\,\,1\,\,-\,\,\exp\Big( -  \bas^2\, Const\,  r^2 ln\Lb r^2 Q^2_s(Y=Y_0)\Rb\Big)\,\,=\,\,
1 - \exp\Big( - r^2 \,Q^2_s\Lb A;Y=Y_0\Rb\Big)
\eeq
 The last equation is a  simplification of the original formula but it reflects the main physics of saturation and considerable simplify calculations.

\eq{ADA1} can be translated into the boundary conditions for $\phi$ on the line $Y=Y_0$ ($\xi_s=0$, see \fig{2}) that has the following form:
\beq \label{BC}
\phi\Lb \xi_s=0; \xi\Rb \,\,=\,\,\phi_0 e^{\xi}
\eeq
while solution of \eq{SOL} gives  quite a different function at $\xi_s$ = 0 (see \fig{incon}). Therefore we need to find a more general solution than it is given by \eq{GSOL}.
\FIGURE[h]{\begin{minipage}{7cm}{
\centerline{\epsfig{file=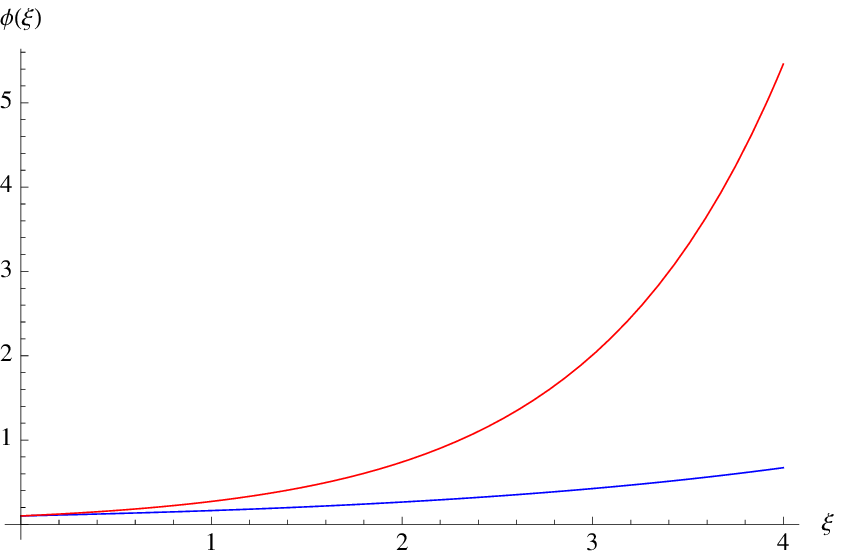,width=65mm}}}
\end{minipage}
\begin{minipage}{6cm}{
\caption{ Boundary conditions at Y=$Y_0$. The red(upper)  line is  $\phi(\xi)$ from McLerran-Venugopalan formula while the blue (lower) one is the solution of \eq{SOL}.}
\label{incon}}
\end{minipage}
}

One of the general features of solution of \eq{SOL} is the increase of $\phi$ in the saturation region. it means that only in the vicinity of the critical line we have to keep term $\exp\Lb - \phi \Rb$. Inside of the saturation region we can neglect this term reducing the equation to the simple one, namely,
\beq \label{EQC4}
\phi_{\xi_s, \xi}\,\,=\,\,\frac{1}{4};\,\,\,\,\,\mbox{or}\,\,\,\,\,\,\frac{\partial^2 \phi}{\partial t^2} \,\,-\,\,\frac{\partial^2 \phi}{\partial x^2}\,\,=\,\,\frac{1}{4}
\eeq
with the initial and boundary conditions of \eq{IC} and \eq{BC}, respectively.

It is well known that the solution of this equation is different for $t = z < x\;( \xi <0)  $ and $t = z > x\;(\xi  > 0 )$\cite{MATH}.  For $ t = z < x \;(\xi <0)$ the solution is not affected by the boundary conditions and it has the form
\beq \label{SOL11}
\phi _1\Lb z \Rb \,\,=\,\,\frac{1}{8} z^2\,\,+\,\,\frac{\phi_0}{2} \,z\,\,+\,\,\phi_0
\eeq
\FIGURE[h]{
\centerline{\epsfig{file=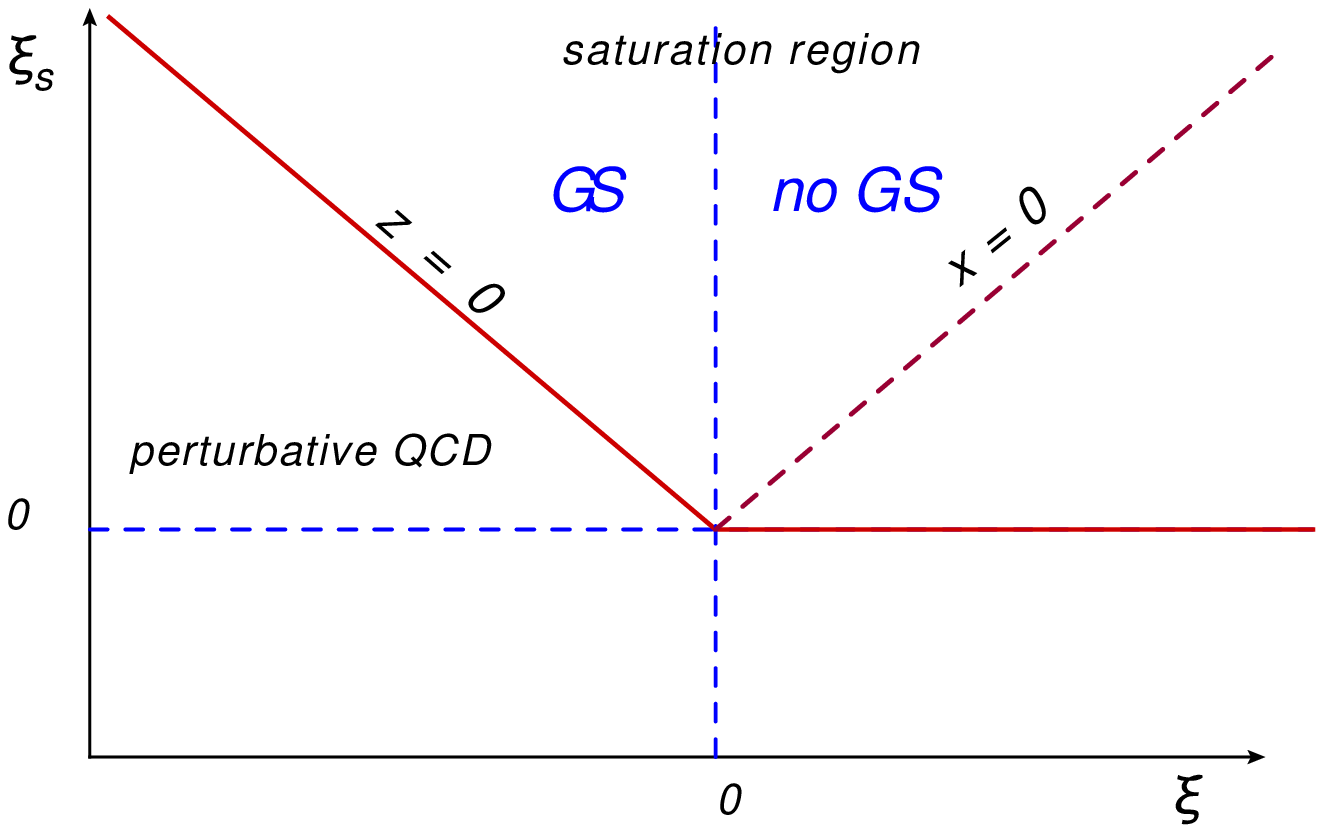,width=120mm}}
\caption{Saturation region: $z = \xi_s + \xi$ and $x = \xi_s - \xi$.}
\label{2}
}
One can see that the general solution to \eq{EQC4} has the form:
\beq \label{GSOL1}
\phi \Lb \xi_s, \xi \Rb \,\,=\,\frac{1}{4}\xi_s\,\xi\, +\, F_1\Lb \xi_s\Rb \,+\,F_2 \Lb \xi\Rb
\eeq
and the solution of \eq{EQC4} can be obtained from \eq{GSOL1} using the restriction from \eq{IC}.
For $t  = z > x\;( \xi > 0 )$ we need to take into account the boundary condition of \eq{BC}. Using the general solution in the form of \eq{GSOL1} and the matching condition on the line $\xi=0$
\newpage
\FIGURE[h]{
\begin{tabular}{c  }
\epsfig{file=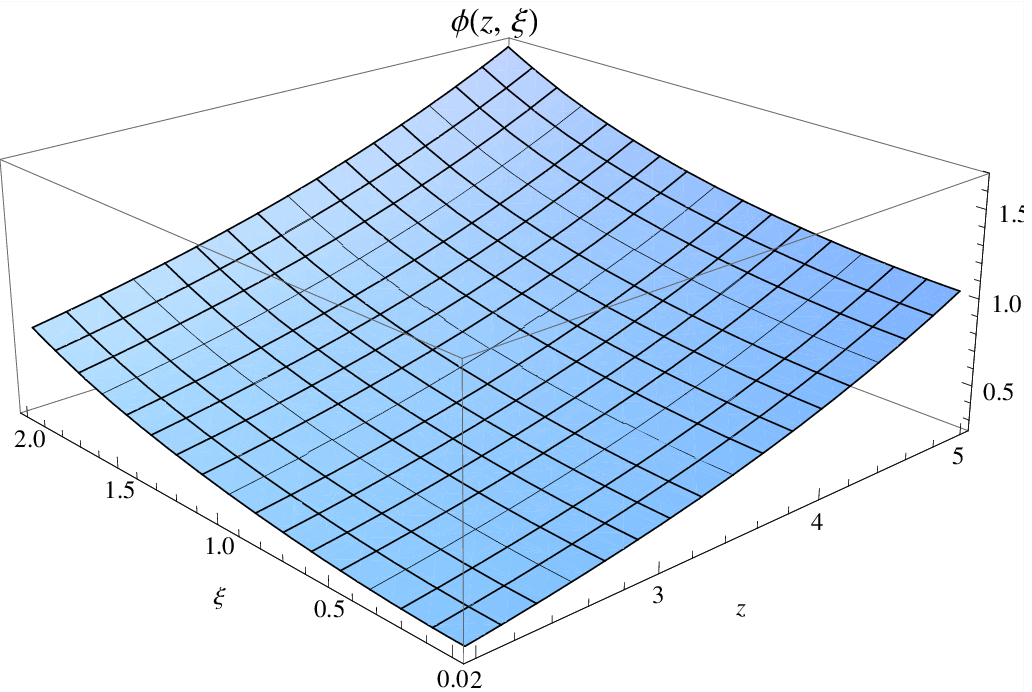,width=70mm}\\
\fig{exact}-a\\
\epsfig{file=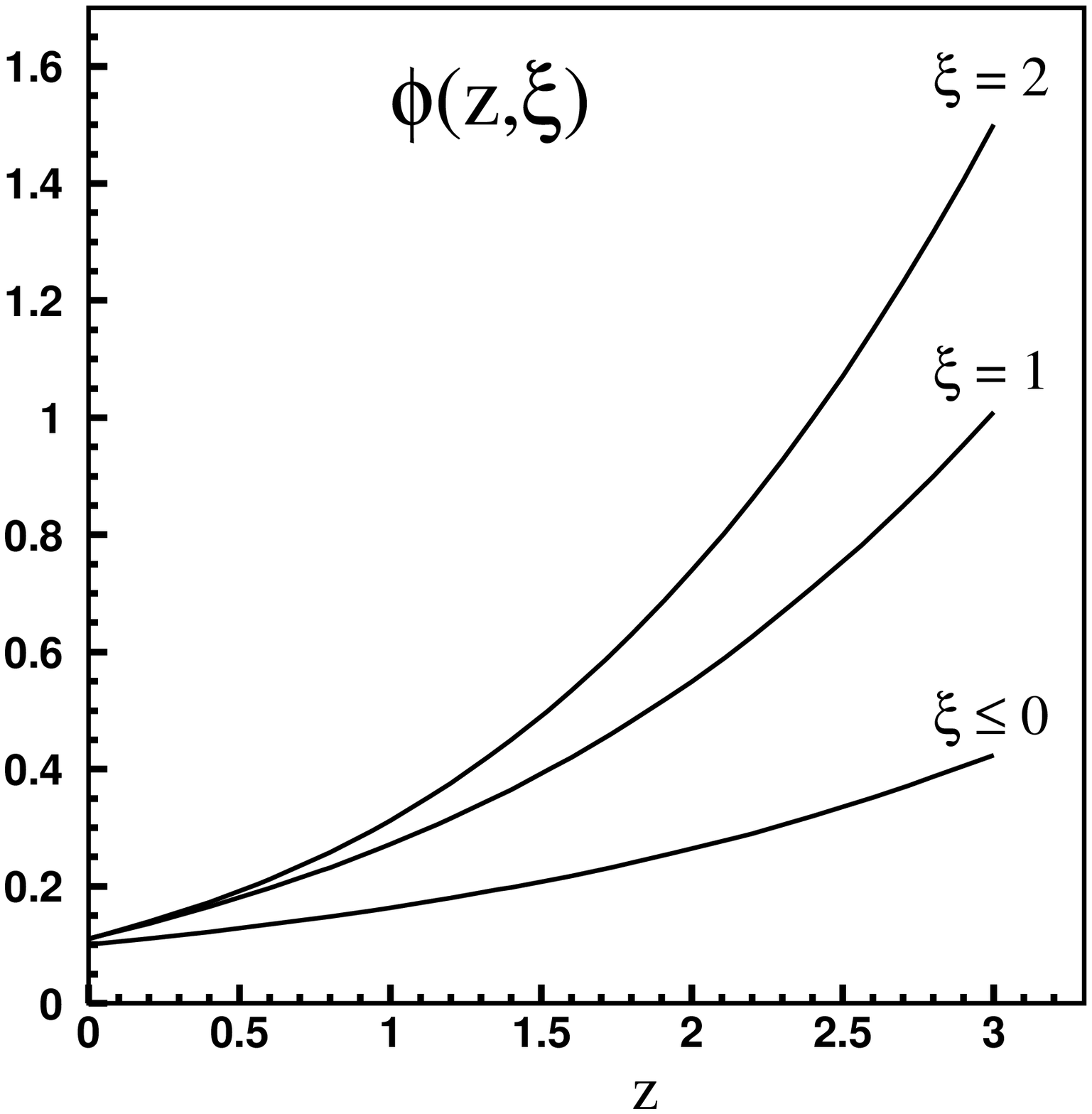,width=70mm}\\
\fig{exact}-b \\
\end{tabular}
\caption{The exact solution of \eq{EQXIY}  for function $\phi\Lb z, y \Rb$(\fig{exact}-a) and the dependence on $z$ at different values of $\xi$ (\fig{exact}-b). $y = \xi_s$,
$ \xi_s = z - \xi $ and $\phi_0$ is taken to be equal to 0.1. }
\label{exact}
}
\beq \label{MC}
\phi_1\Lb \xi = 0\Rb \,=\,\phi_2 \Lb\xi =0\Rb
\eeq
simultaneously with the boundary conditions that has the form
\beq \label{BC1}
\phi_2 \Lb\xi_s = 0 \Rb\,\,=\,\,\phi_0 e^{\xi}
\eeq
we obtain the following solution for $\xi > 0 $
\beq \label{GSOL2}
\phi_2\Lb z, \xi\Rb\,\,=z^2/8\,-\,\xi^2/8 \,+\,\phi_0\,e^{\xi}\,+\,\h \phi_0 \,\xi_s
\eeq
Therefore, the solution to the simplified \eq{EQC4} has the following form
\bea \label{SOLEQC}
\phi\Lb \Rb\,\,=\,\,\left\{\begin{array}{l}\,\,\,\phi_1\Lb z \Rb\,\,\,\,\,\mbox{for}\,\,\,\xi\leq \,0\,;\\ \\
\,\,\,\phi_2\Lb z, \xi\Rb
\,\,\,\,\,\mbox{for}\,\,\,\xi\,>\,0\,; \end{array}
\right.
\eea
For the solution of the general \eq{EQXIXIS} we have
\bea \label{SOLEQXIXIS}
\phi\Lb \xi_s,\, \xi;\, \Rb\,\,=\,\,\left\{\begin{array}{l}\,\,\,\phi\Lb z ; \,\eq{SOL}  \Rb\,\,\,\,\,\mbox{for}\,\,\,\xi\leq \,0\,;\\ \\
\,\,z^2/8\,-\,\xi^2/8 \,+ \,\phi\Lb \xi_s; \,\eq{SOL}\Rb \,-\,\phi_0\,+\,\phi_0\,e^{\xi}\,+\,\h \phi_0 \,\xi_s
\,\,\,\,\,\mbox{for}\,\,\,\xi\,>\,0\,; \end{array}
\right.
\eea
In \eq{SOLEQC} we assumed that for $\xi > 0$ we are approaching the solution of \eq{EQC4}.

One can see that solution of \eq{SOLEQXIXIS} does not show the geometric scaling behavior and solution of \eq{EQXIXIS} depends both on  $z$ and $\xi$.
It happens so due to the influence of the boundary conditions.

In \fig{exact} we plot the numerical solution\footnote{It is worth to mention that \eq{EQXIXIS} has the form which does not depend on extra parameters and, using the numerical solution, we do not loose the generality of our approach.} \,of  \eq{EQXIXIS} in the region
$\xi \,>\,0$ (see \fig{exact}) with the following boundary conditions:
\bea \label{BIC}
\phi\Lb y \equiv  \xi_s,\xi = 0\Rb\,\,&=&\,\,\phi\Lb \xi_s\Rb;\nn\\
\phi\Lb y \equiv  \xi_s\,=\,0,\xi \Rb\,\,&=&\,\,\phi_0\,e^{\xi};
\eea
One can see that this solution does not show the geometric scaling behavior inside the saturation   domain.

To preserve the geometric scaling behavior we need to assume  that for $\xi > 0$ at $Y=Y_0$ the behavior of the scattering amplitude is not given by the Glauber\\(McLerran-Venugopalan) formula but rather is determined by the solution of \eq{SOL}.

The initial condition based on McLerran-Venugopalan formula stems from the main assumption that there exists the rather low energy at which dipole rescatters in the nucleus but the emission of gluons can be neglected. At first sight,  we do have  arguments why the emission is small.
Indeed,  if $\bas^2 A^{1/3} \approx 1$ for $\bas Y  <1$ (or $Y \ll 1/\bas \approx 1/A^{1/6}$)  the emission of the gluon will be suppressed while the Glauber-type rescattering will be essential since the interaction with the nucleus will be proportional to $\bas A^{1/3} \approx 1$.  Therefore, we can choose the energy ($x$) which is large enough to use only the exchange of gluon for the dipole amplitude while the emission of gluon will be still suppressed.  For this kinematic region we showed that the geometric scaling behavior of the amplitude is not valid. Since the McLerran-Venugopalan
formula follows from the CGC approach and represents its key feature, we can claim that the CGC leads to the violation of the geometric scaling behavior
in the kinematic region where $r^2\,Q_s\Lb Y = Y_0 \Rb \,>\,1$.

However  high density QCD  has two facets at the moment:  the color glass condensate {CGC} approach \cite{MV,JIMWLK,B,KOVLU}  and  the BFKL Pomeron calculus\cite{BFKL,GLR,MUQI,BRN,BART}. Both these approaches lead to the same non-linear Balitsky-Kovchegov  equation \cite{B,K} for the dilute-dense system scattering in the large $N_c$  approximation which we consider here. However, in the BFKL Pomeron calculus the emission of gluons is taken into account  even at small values of energy. In this approach the natural initial condition is $N^{eff}_N\Lb r,Y = Y_0 \Rb\,\,=\,\,r^2\,Q_s(N;Y_0) $ (compare with \eq{ADA1}) and the value of $Y_0$ is much smaller that $Y_A\,=\,(1/3)\ln A$. This case we  consider in the next section.

Before doing this we would like to draw your attention to the fact that the condition $\bas^2 A^{1/3} \approx 1$ can be reached in QCD only for scattering of states with typical extremely short distances (say, onium which made of two very heavy quarks). Running QCD coupling  for such states could be small of the order of $ \bas \sim  1/A^{1/6}	$. For nuclei the typical $\bas$ is determined by the size of nucleons and could be as small as $\bas = 0.2 \div 0.3$ but not smaller. In this case the situation changes crucially:
 summing all powers of $\bas Y$ will lead us to the BFKL contribution namely $\bas^2 e^{\omega_0 Y}$.
This contribution can be larger than the re-scattering in the classical gluon fields. It happens so
 at  large  $Y_A = \ln A^{1/3}$ since  $ \omega_0$ is larger that $1/3$ for $\bas = 0.2 \div 0.3$ (see \eq{KDA1}).

\section{Solution for two critical lines}
\setcounter{equation}{0}


\begin{boldmath}
\subsection{Equation for $ Y < Y_A$}
\end{boldmath}


In the framework of the BFKL Pomeron calculus the rescattering with large rapidities but smaller than $Y_A = \ln A^{1/3}$ should be treated using the non linear equation.
In this kinematic region each dipole interacts with the number of nucleons that are smaller than $\rho\,R_A \sim A^{1/3}$ and which actually is equal to $e^{Y}\rho/m $ where $\rho$ is the density of nucleons in the nucleus and  $m$ is the proton mass\cite{LERYA,QIU}.  We can  incorporate this observation into \eq{MEQ5} by changing the definition of $T_A\Lb b \Rb$ in \eq{MEQ3}, namely,
\beq \label{TTT}
T_A\Lb b;Y\Rb\,\,=\,\,\int^{+ 1/mx}_{-1/mx}  d z\, \rho\Lb b, z\Rb
\eeq
For $x \ll \, x_A = \,e^{- Y_A}$ \eq{TTT} reduces to \eq{MEQ3} while for $1 \,\gg \,x \,\gg \,x_A$ \eq{TTT} leads to
$e^{Y}\rho/m $ for cylindrical nuclei.

Introducing
\beq \label{NAN}
N_A\Lb Y, \xi\Rb\,\,=\,\,T_A\Lb b; Y\Rb\,N^{eff}_N\Lb Y, \xi\Rb
\eeq
in stead of \eq{NASAT} and using $T_A\Lb b; Y\Rb$ in the form:
\bea \label{CYA}
T_A\Lb b; Y\Rb\,\,=\,\,\left\{\begin{array}{l}\,\,\,\rho\,2 R_A \,\propto\,A^{1/3}\,\,\,\,\,\mbox{for}\,\,\, Y \leq \,Y_A\,;\\ \\
\,\,\,e^{Y}\rho/m
\,\,\,\,\,\mbox{for}\,\,Y\,<\,Y_A\,; \end{array}
\right.
\eea
we can re-write \eq{MEQ5} in the following form:
\bea
x \,<\,x_A \,\,(Y > Y_A)\,
\,&:& \frac{d N_A\Big(Y; \xi \Big)}{d Y}\,\,=\,\,\bas \big\{ \int^\xi_{\xi_s}\,d \xi' N_A \Big(Y; \xi' \Big) \,\,-\,\,N_A^2\Big(Y; \xi \Big) \Big\}; \label{EQ1}\\
x \,>\,x_A \,\,(Y < Y_A)\,
\,&:& \frac{d N_A\Big(Y; \xi \Big)}{ d Y}\,-\,N_A\Big(Y; \xi \Big)\,\,=\,\,\bas \big\{ \int^\xi_{\xi_s}\,d \xi' N_A \Big(Y; \xi' \Big) \,\,-\,\,N_A^2\Big(Y; \xi \Big)\Big\} ;
\label{EQ2}
\eea

We will solve these two equations and these solutions should be matched on the line $Y = Y_A$ (see \fig{1}).
These two equations have different critical lines. The critical line of the first one (see \eq{EQ1})   has been discussed in \eq{TR1} and \eq{GACREQ}. It is shown as line 2 in \fig{1} and has the form
\beq \label{CRL2}
\xi\,\,\equiv\,\,\ln\Big( r^2\,Q^2_s\Lb A; Y_A\Rb\Big)\,\,=\,\,-\,\xi_{2 s}\,\,=\,\,-\,4\,\bas\,\Lb Y \,-\,Y_A\Rb
\eeq
The easiest way to find the critical line for \eq{EQ2} is to search the solution to the general \eq{MEQ5} with $T_A\Lb b \Rb$ replaced by $T_A\Lb b; Y\Rb$ in the semi-classical form
\beq \label{SCSOL}
N_A\Lb Y, \xi\Rb\,\,= \,\,e^{S\Lb Y, \xi\Rb}\,\,=\,\,e^{\omega\Lb Y,\xi\Rb\,Y\,-\,\Lb 1 - \gamma\Lb Y; \xi\Rb\Rb\,\xi\,+\,S_0}
\eeq

This solution has a form of wave-package and the critical line is the specific trajectory for this wave-package which coincides with the its front line. In other words, it is the  trajectory on which the phase velocity ($v_{ph}$) for the wave-package is the same as
the group velocity ( $v_{gr}$). The equation $v_{gr} \,\,=\, v_{ph}$ has the following form fort \eq{EQ2}
\beq \label{CRL1}
v_{ph}\,\,=\,\,\bas \frac{\chi\Lb \gamma_{cr} \Rb}{ 1 \,-\,\gamma_{cr}}\,+\,\bas \frac{1}{ 1 \,-\,\gamma_{cr}}\,\,=\,\,- \bas \chi'\Lb \gamma_{cr}\Rb\,\,=\,\,v_{gr}
\eeq
Solution to \eq{CRL1} gives $\gamma_{cr}\,\,=\,\,\sqrt{\bas}\,+\,{\cal O}(\bas)$ and it leads to the equation (see \fig{1})
\beq \label{CRL12}
\xi\,\,=\,\,-\xi_{1s}\,\,=\,\,\Lb 1\,+\,2\,\sqrt{\bas}\Rb \Lb Y_A - Y\Rb
\eeq

\FIGURE[h]{
\centerline{\epsfig{file=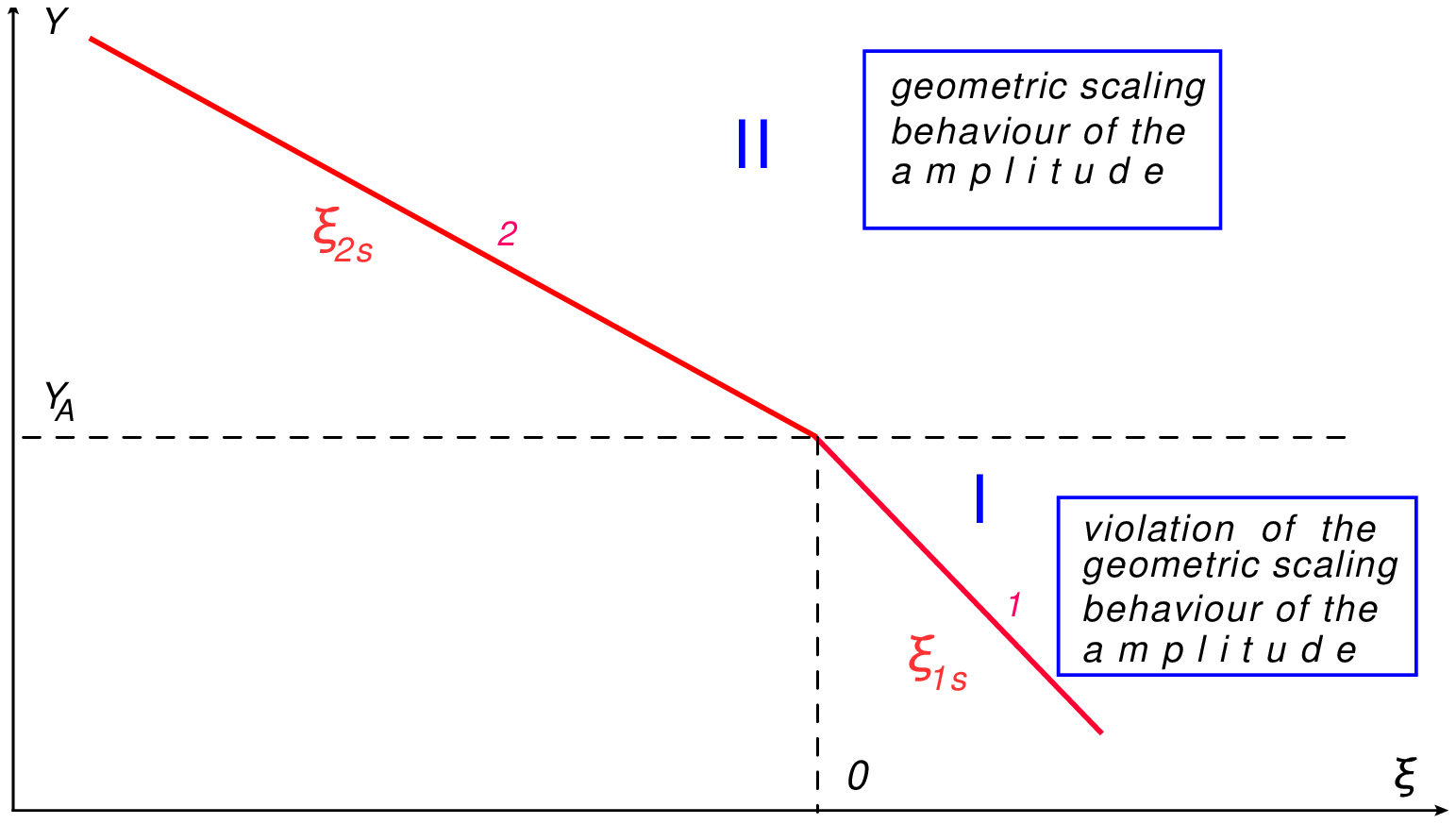,width=120mm}}
\caption{Two  saturation regions for $N$: region I for $Y < Y_A= (1/3)\,\ln A$ and
$\xi > \xi_{1s} = ( 1 + 2 \sqrt{\bas}) ( Y_A- Y) $; and region II  $Y > Y_A= (1/3),\ln A$ and
$\xi \,>\,\xi_{2s} = -4 \bas (Y - Y_A) $. Here $\xi = \ln \Lb r^2\,Q_s(Y=Y_0; A)\Rb$ where $r$ is the dipole size}
\label{11}
}

\subsection{Solutions}
In both regions we will look for solutions using
\beq \label{NS}
\widetilde{N}_A\Lb Y; \xi\Rb\,\,=\,\,\int^{\xi}_{\xi_{i s}}
 d \xi'\,\Big( 1\,-\,e^{ - \phi\Lb Y; \xi \Rb}\Big)
\eeq
In region II we obtain \eq{EQXIY} which we need to solve with the same initial condition as in \eq{IC}. As we have discussed \eq{SOL} gives the solution of this problem.

In the region I using \eq{NS} we obtain after differentiation over $\xi$
\beq \label{F5}
\widetilde{N}_{A, Y \xi} \,-\,\widetilde{N}'_{A, \xi}\,=\,\,\bas \widetilde{N}_A \Lb 1 \,-\,\widetilde{N}'_{A, \xi}\Rb\;\;\mbox{or}\;\;
\phi_Y\,e^{- \phi} \,-\,\Lb 1 - e^{-\phi}\Rb\,= \,\bas N_A e^{- \phi}
\eeq
Differentiating \eq{F5} over $\xi$ we get
\beq  \label{F6}
\phi_{Y \xi} - \phi_Y \,\phi_\xi -  \phi_\xi\,=\,-\bas  \widetilde{N}_A\phi_\xi + \bas\,\Lb 1 - e^{- \phi}\Rb
\eeq

\eq{F6} has a simple solution for large $Y$ and $\xi$. Indeed, assuming that $\phi $ is large in this region , \eq{F6}
can be re-written in the form
\beq \label{F61}
\phi_{Y z_1}  \,- \,\phi_{z_1 z_1}\,- \,\phi_{Y} \phi_{z_1} \,-\,\phi^2_{z_1}\, - \, \phi_{z_1}\,=\,-\bas z_1 \phi_{z_1}\, +\, \bas\,
\eeq
with\footnote{ For the sake of simplicity we consider $ 2 \sqrt{\bas} \,\ll\,1$ in this expression.} $z_1\,=\,\,\xi + \xi_{1,s}\,=\,\Lb 1 + 2 \sqrt{\bas} \Rb \Lb Y_A - Y\Rb + \xi$.

One can see that the common solution of the two following equations
\beq \label{F7}
1.\,\,\,\,\phi_{Y \xi}\,\,=\,\, \bas; \,\,\,\,\,\,\,\,\,2. \,\,\,\,\,\phi_Y  + 1 \,=\,-\bas \,z_1;
\eeq
will be the solution of \eq{F61}.  It is easily seen that such a solution has the general form
\beq \label{F8}
\phi_{R1}\,\,=\,\ \bas \,\xi \,\Lb Y_A - Y\Rb  \,+\,\bas (Y_A - Y)^2/2\,+\,\Lb Y - Y_A\Rb  \,+\,F\Lb \xi \Rb
\eeq
where $F\Lb \xi\Rb$ is the arbitrary function.  The initial condition for this equation follows from the solution  of \eq{TR3}
where $T_A\Lb b \Rb$ is replaced by $T_A\Lb b; Y \Rb$. They have the form
\beq \label{IC1}
\phi_{R1}|_{\xi = \xi_{1s}} \,=\,\phi_0 \,;\,\,\,\,\frac{d\phi_{R1}}{d\xi}|_{\xi = \xi_{1s}} \, =   \,\phi_0
\eeq

The boundary condition stems from the solution in region II (see \eq{SOL}) and has the form
\beq \label{BC11}
\phi_{R1}\Lb Y = Y_A; \xi \Rb \,\,=\,\,\phi (\xi)
\eeq

Choosing $F\Lb \xi \Rb = \phi\Lb \xi\Rb$ we see that  solution $\phi_{R1}$   matchers   the boundary condition of \eq{BC11} but not the initial condition of \eq{IC1}.  We need to solve \eq{F6}  in the region of small $z_1$ to satisfy this condition, but in this region we cannot neglect  the term $\exp\Lb - \phi \Rb$ in \eq{F6}.
We can approach this region  solving
\eq{F5} which can be rewritten in the form:
\beq \label{F51}
\phi_Y\Lb Y, z_1\Rb - e^{\phi\Lb Y, z_1\Rb}\,+\,1\,\,=\,\,\bas \widetilde{N}_A\Lb Y; z_1\Rb
\eeq
with
\beq \label{NT}
\widetilde{N}_A\Lb Y; z_1\Rb\,\,=\,\,\int^{z_1}_{0}
 d z'_1\,\Big( 1\,-\,e^{ - \phi\Lb Y; z'_1 \Rb}\Big)
\eeq

After differentiation of \eq{F51} with respect to $z_1$ one obtains
\beq \label{F52}
\phi_{Y,z_1}\Lb Y, z_1\Rb \,-\,\phi_{z_1}\Lb Y, z_1\Rb\, e^{\phi\Lb Y, z_1\Rb}\,\,=\,\,\bas \Big( 1\,-\,e^{-\phi\Lb Y, z_1\Rb}\Big)
\eeq

The initial and boundary conditions for \eq{F52} looks as follows:
\bea\label{INBC}
\mbox{initial conditions:} & ~~~~&\phi\Lb Y, z_1=0\Rb\,=\,\phi_0\,;\nn\\
\mbox{boundary  conditions:} & ~~~~&\phi\Lb Y=0, z_1\Rb\,=\,\phi\Lb z'_1\Rb\,;
\eea

This equation has been solved numerically and the solution for $\phi$ and $N_A = 1 - \exp\Big( - \phi \Big)$ is shown in     \fig{solR1}.


\FIGURE[h]{
\begin{tabular}{c c }
\epsfig{file=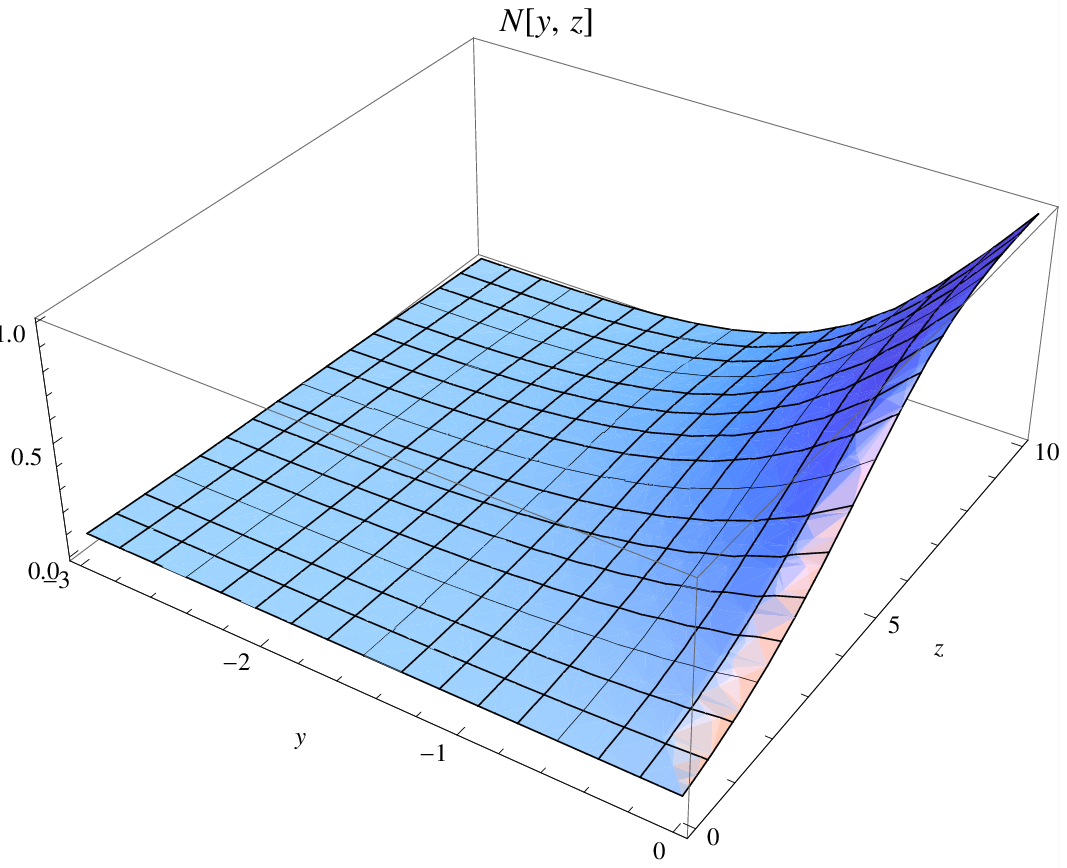,width=70mm} &\epsfig{file=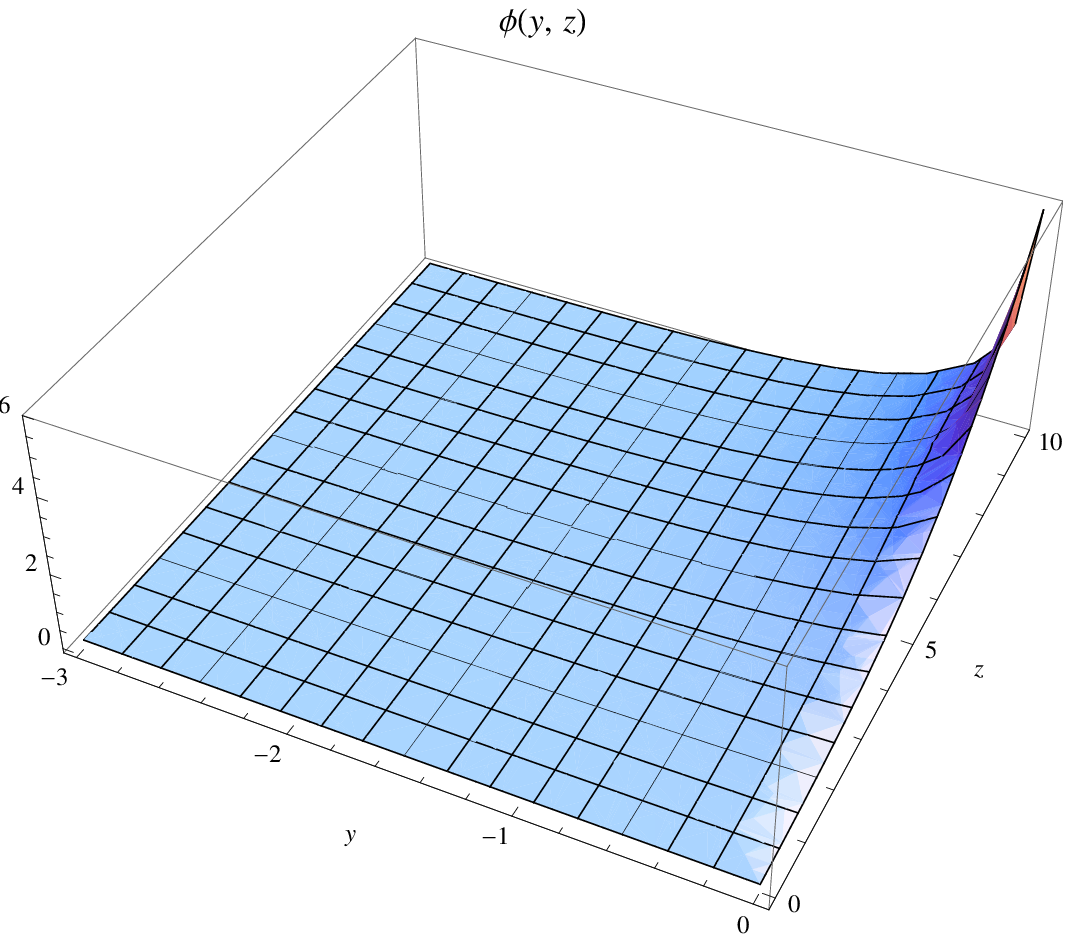,width=70mm}\\
\fig{solR1}-a & \fig{solR1}-b  \\
\end{tabular}
\caption{The exact solution of \eq{F52}  for function $N_A\Lb y, z\Rb\,\equiv\,N\Lb y,z\Rb$ (\fig{solR1}-a)  and  $\phi$ (\fig{solR1}-b). $y = \xi_{1s}$,
$ \xi_{1s} = z_1 - \xi $ and $\phi_0$ is taken to be equal to 0.1 and $\bas = 0.2$. }
\label{solR1}
}


Therefore, the full solution has the geometric scaling behavior in the region II but shows the violation of the scaling behavior in region I as it follows from \eq{F8} and \fig{solR1}.

Comparing this result with the conclusions of the previous section we see that the BFKL Pomeron calculus predicts the geometric scaling behavior in the saturation region for $r^2\,Q^2_s\Lb A; Y_0\Rb\,>\,1$ and $Y >Y_A$.

\section{Impact parameter dependence of the scattering amplitude}
In this section  we complete the study of the impact parameter dependence of the scattering amplitude that has been started in Ref.\cite{KOLE}. In Ref.\cite{KOLE}  we claim that in the framework of the BK equation with the simplified kernel the impact parameter dependence can be absorbed into redefinition of the  saturation scale, namely,
\beq \label{BD1}
\mbox{For  proton}\,\,\,\tau\,\equiv\,r^2\, Q^2_s(P; x)\,\,\,\longrightarrow
\,\,\,\,r^2\, Q_s(A; x)\;\;\mbox{with}\;\;Q_s\Lb A; x \Rb \,=\,T^2_A\Lb b \Rb\, Q^2_s\Lb P; x\Rb \,\,\equiv\,\,\tau_A\,\,\,\,\mbox{for nuclei}
\eeq
 or
\beq \label{BD2}
z_{\mbox{proton}}\,\,\,\longrightarrow \,\,\,\,z_{\mbox{nucleus}}\,\,=\,\,z_{\mbox{proton}}\,\,\,+\,\,\,2\,\ln\Big( T_A\Lb b \Rb\Big)
\eeq

\FIGURE[t,h]{
\begin{tabular}{c c }
\epsfig{file=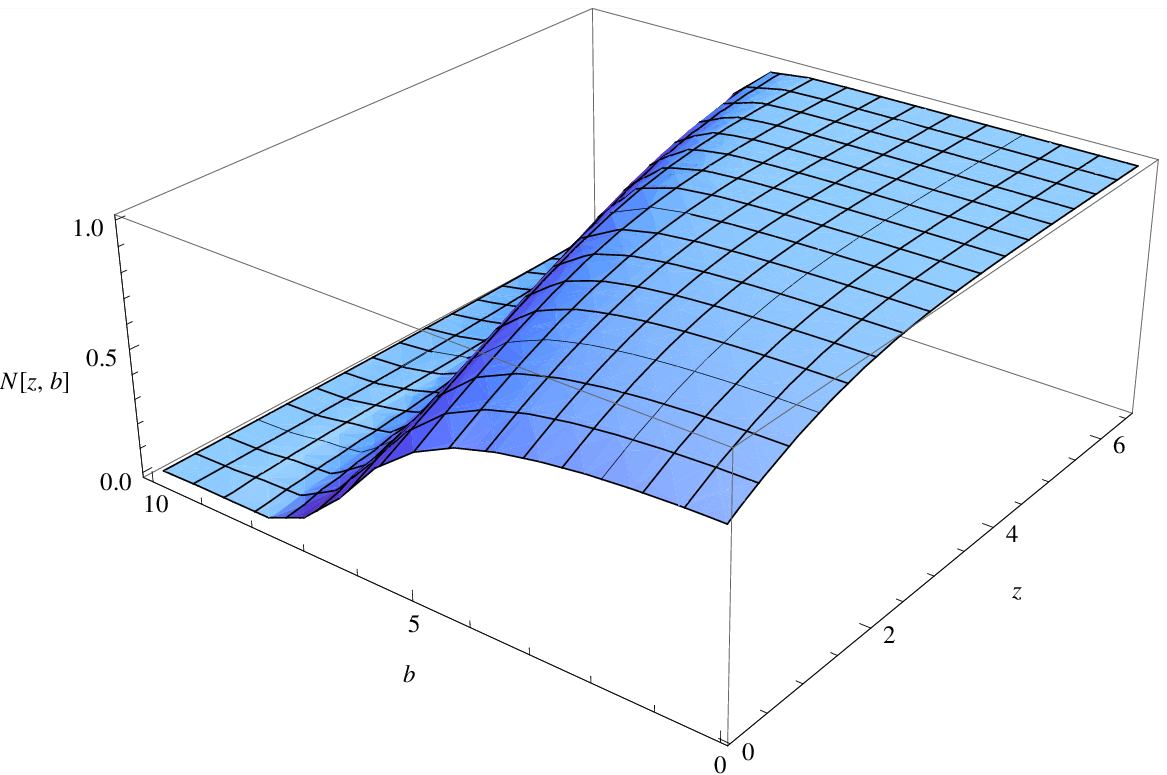,width=85mm} &
\epsfig{file=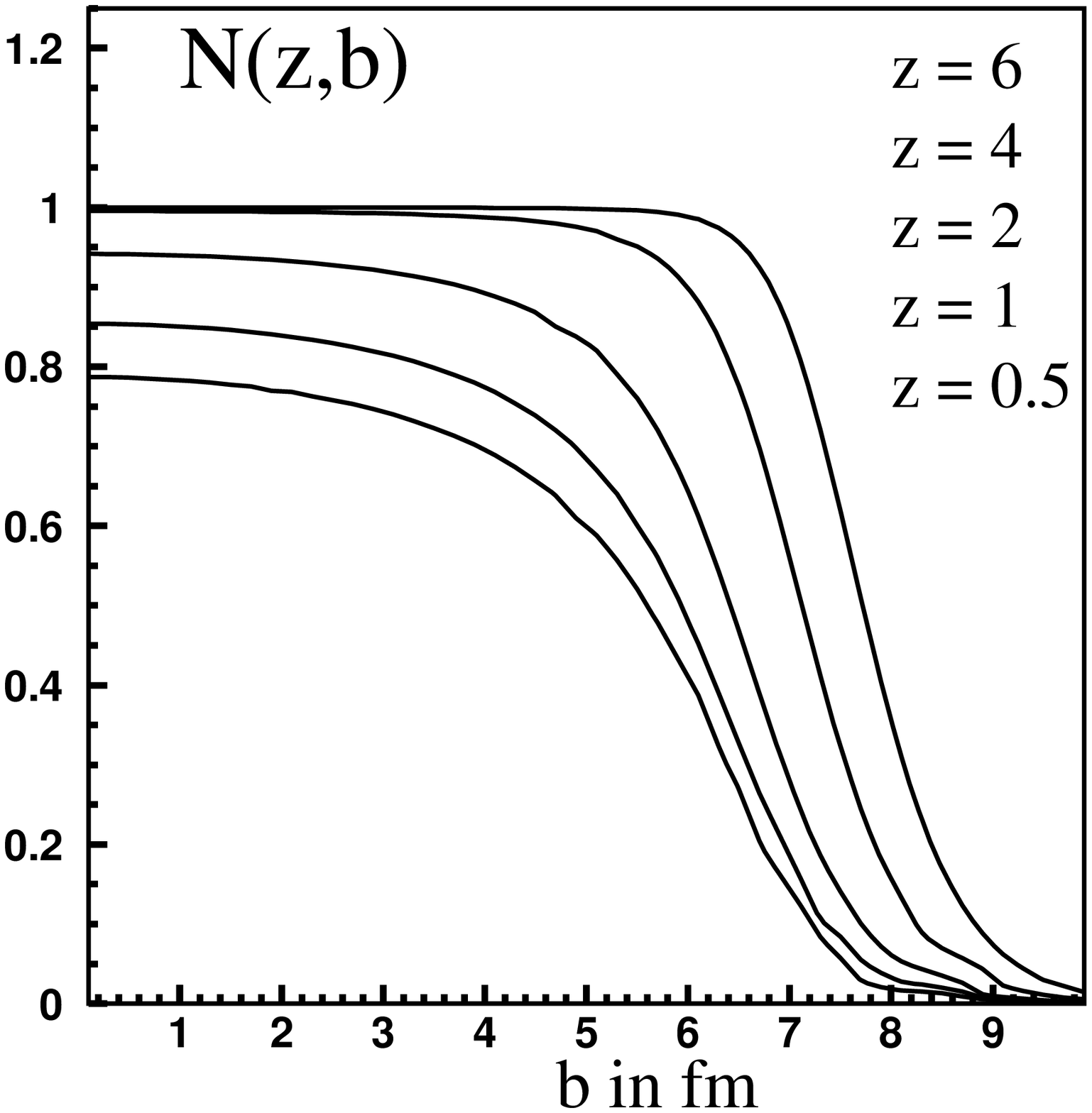,width=65mm}\\
\fig{bdep}-a &\fig{bdep}-b\\
\end{tabular}
\caption{The exact solution of \eq{SOLL}  for function $N\Lb z,b\Rb= 1 - \exp\Lb- \phi\Lb z, b \Rb\Rb$ (\fig{bdep}-a) for the interaction of the dipole with the gold  and its dependence on $b$ at different values of $z$ (\fig{bdep}-b). $y = \xi_s$,
$ \xi_s = z - \xi $ and $\phi_0$ is taken to be equal to 0.1. }
\label{bdep}
}

However, this claim is based on the solution of \eq{SOL} which assumed that $\phi_0 \,\ll\,1$. In the general solution of \eq{SOLL} one can see that $b$-dependence cannot be reduced to changes in the value of the saturation scale. In \fig{bdep} we plot the dependence of the scattering amplitude on the impact parameter using the realistic $T_A\Lb b \Rb$ for the gold\cite{WS} and the value of $\phi_0$ in \eq{TR3} taken from the fit of the HERA data \cite{WK}.
The first glance at \fig{bdep} shows that the typical value of $b$ increases with $z$. It has a natural explanation. Indeed the width $b_o$ of $b$ distribution we can define as $N\Lb z, b=b_0\Rb = e^{-1}$.  Since the amplitude $N$ has the geometric scaling behavior the value of $b_0$ can be determined from the equation
\beq \label{BD3}
N_A\Lb \tau_A\Rb \,\,=\,\,N_A\Big(T^2_A\Lb b_0 \Rb\, Q^2_s\Lb P; x\Rb\Big)\,\,=\,\,e^{-1}\,\,
\eeq
where $Q_s\Lb P; x\Rb$ is the saturation scale for the proton target.
In other words, the typical $b$ can be determined   from the following equation:
\beq
\label{BD4}
 T^2_A\Lb b_0 \Rb\, Q^2_s\Lb P; x\Rb\,\,\,=\,\,\tau_A \,\,=\,\,\,\mbox{Const}
\eeq

Recalling that
\beq \label{BD5}
Q^2_s\Lb P; x\Rb\,\,=\,\,Q^2_0\Big(\frac{1}{x}\Big)^\lambda\,\,\,\,\,\,\mbox{with}\,\,\,\,\,\,\,\,\lambda\,=\,4 \bas
\eeq
one can see that
\beq \label{BD6}
b_0\,=\,\,R_A \,\,+\,\,\h\,h\,\lambda \ln(1/x)
\eeq
where we use that
\beq \label{BD7}
T_A\Lb b_0 \Rb\,\,\,\,\,\,\xrightarrow{ b > R_A}\,\,\,\,\,\,e^{ - \frac{b - R_A}{h}}
\eeq
and $h $ is about  $0.5 \,fm$.

In terms of $z$  \eq{BD7} has even a simpler form:
\beq \label{BD8}
b\,\,=\,\,R_A\,+ \frac{h}{2}\,z_{\mbox{proton}}
\eeq
It is wort mentioning that \eq{BD8} does not depend on the specific form of energy dependence of the saturation momentum.

It should be stressed that  we obtain  a logarithmic increase of the radius of interaction with $z$. It stems from \eq{MEQ5} where we have integrated over the impact parameter of the nucleon. As we have discuss we can trust this $b$-dependence in the kinematic region of \eq{MEQ6}. Certainly, for the dipole-gold scattering for $z \leq 7$ we can use this Glauber-type approximation. All problems with $b$-dependence are originated from the large $b$ dependence of the dipole-proton amplitude which falls down as $1/b^4$ in perturbative QCD\cite{KOWI}. Implicitly we assumed that the non-perturbative corrections to the B-K equation has been taken into account for dipole-nucleon scattering in the transition  from \eq{MEQ1} to  \eq{MEQ5}.
\begin{figure}
\begin{minipage}{80mm}{
\centerline{\epsfig{file=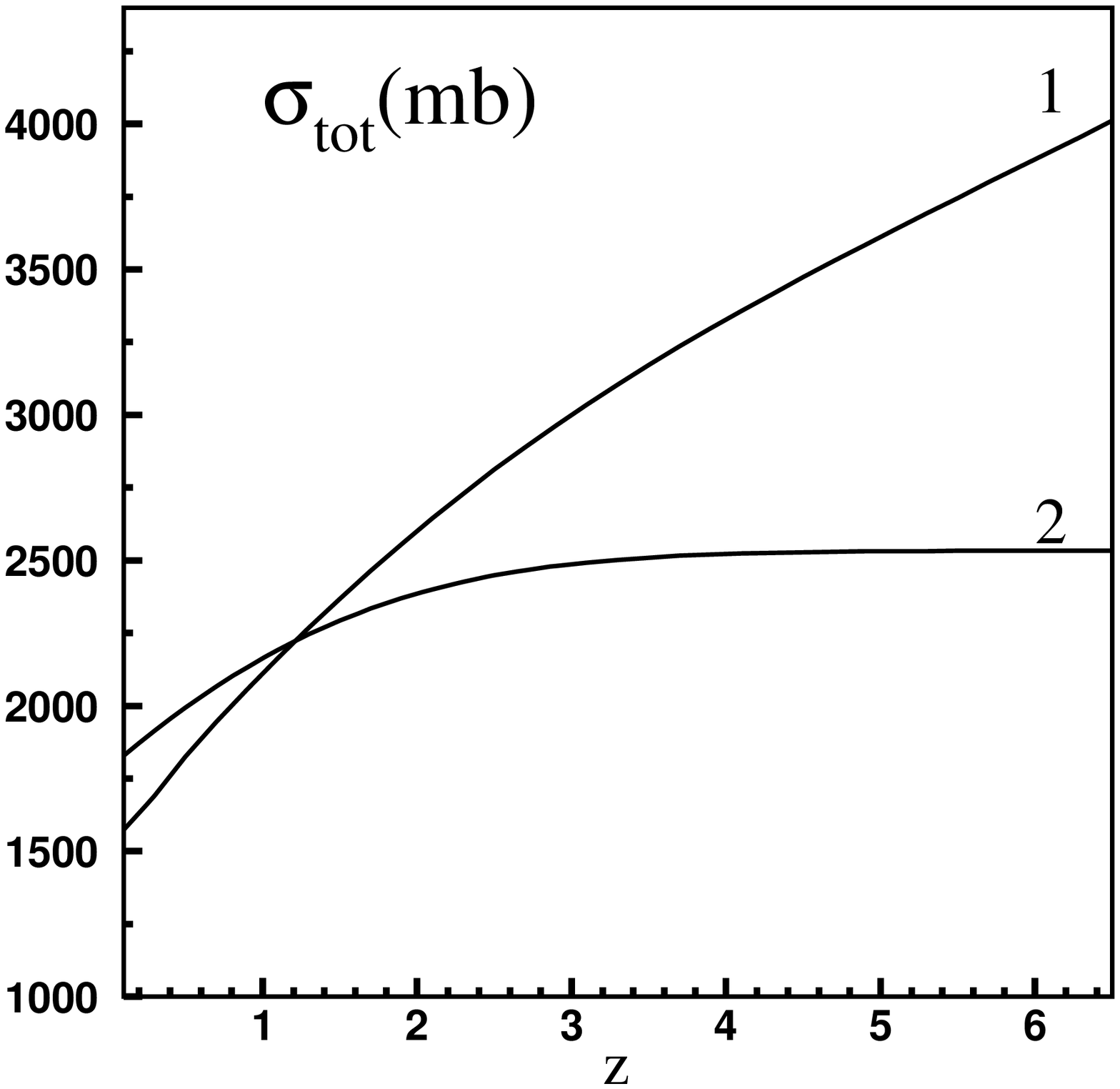,width=70mm}}
}
\end{minipage}
\begin{minipage}{70mm}{
\caption{ Total cross section for dipole-gold amplitude $\sigma_{tot}= 2 \int d^2 b\, N\Lb z,b\Rb$ in the saturation region: the solution of \protect\eq{SOLL} (curve 1) and the solution to \protect\eq{EQXIXIS} with simplified $T_A\Lb b \Rb \,\varpropto\, \Theta\Lb R_A - b\Rb$ (curve 2).}
\label{sigmab}
}
\end{minipage}
\end{figure}
In \fig{sigmab} we plotted the dependence on $z$ the total cross section of dipole-nucleus intertaction for the case of gold
\beq \label{BD81}
\sigma_{tot}\,\,=\,\,2 \int d^2b\, N\Lb z,b \Rb
\eeq
One can see a significant difference between realistic Wood-Saxon $T_A\Lb b \Rb$ and the simplified one $T_A\Lb b\Rb\,  = \,\rho\, \Theta \Lb R_A - b\Rb$ where $\rho$ is the density of the nucleons in the  nucleus. This picture demonstrates the significance  of correct $b$-dependence for calculation of the physical observables.

\section{Conclusions}
We hope that we answered three questions that we have discussed in the  introduction. The first one:
 could the initial conditions affect the behavior of the scattering amplitude at $\tau \,\gg\,1$.
The answer is yes and we gave the explicit solution of Balitsky-Kovchegov equation which shows the violation of the geometric scaling behavior of the scattering amplitude if you use the McLerran\,\,-\,\,Venugopalan formula as the initial condition .

The second question: can we trust the McLerran-Venugopalan formula  deeply in the saturation region, has a kind of negative answer. In the sense that  the McLerran-Venugopalan formula cannot be considered as the unique initial condition. We demonstrated that in the BFKL Pomeron calculus  this formula should be replaced by  the solution to the non linear equation in the region of $Y < Y_A = \ln(A^{1/3})$ for very heavy nuclei. This statement gives the answer to the third question: what initial condition we need to use to reproduce the geometric scaling behavior.

It is well known that we have two approaches to high density QCD; the BFKL Pomeron calculus\cite{GLR,MUQI,BRN,BART} and Color Glass Condensate\cite{MV,JIMWLK,B,K}. Both lead to  the same Balitsky-Kovchegov equation for DIS. The difference between them lays
in the initial conditions. For the CGC the initial condition is the McLerran-Venugopalan formula which is valid in the classical gluon field approximation. On the other hand, for the BFKL Pomeron calculus  a natural initial condition stems from the solution of B-K equation  for $Y < Y_A$. Therefore, we can
formulate the main result of the paper in the following way.
The CGC approach leads to the violation of the geometric scaling behavior for DIS with heavy nuclei for $r^2\,Q_s\Lb A; Y =Y_0\Rb \,>\,1$ while the BFKL Pomeron calculus leads to the geometrical scaling behavior of the amplitude for $Y> Y_A$. This result gives a possibility to check experimentally which of these two approaches is more adequate.

\section*{Acknowledgements}

This work was supported in part by the  Fondecyt (Chile) grant 1100648.



\begin{thebibliography}{99}
\bibitem{BALE}
J.~Bartels, E.~Levin,
  Nucl.\ Phys.\  {\bf B387 } (1992)  617-637.
\bibitem{GLR}
L. V. Gribov, E. M. Levin and M. G. Ryskin, {\it Phys. Rep.}\,
{\bf 100}, 1 (1983).

\bibitem{MUQI}
A. H. Mueller and J. Qiu,  {\it Nucl. Phys.},427 {\bf B 268}
(1986) .


\bibitem{MV}
L. McLerran and R. Venugopalan, {\it  Phys. Rev.}  {\bf D 49},2233,
3352  (1994); {\bf D 50},2225 (1994); {\bf D 53},458 (1996); {\bf
D 59},09400
(1999).
\bibitem{JIMWLK}
~J.~Jalilian-Marian, A.~Kovner, A.~Leonidov and H.~Weigert,
{\it  Phys.\ Rev.}\,  {\bf D59}, 014014 (1999),
[arXiv:hep-ph/9706377];\,\,  {\it Nucl.\ Phys.}\,{\bf B504}, 415
(1997),
[arXiv:hep-ph/9701284]; \,\,\,
J.~Jalilian-Marian, A.~Kovner and H.~Weigert,
  {\it Phys.\ Rev.}  {\bf D59}, 014015 (1999),
  [arXiv:hep-ph/9709432];\,\,\,
 A.~Kovner, J.~G.~Milhano and H.~Weigert,
 {\it  Phys.\ Rev.}  {\bf D62}, 114005 (2000),
  [arXiv:hep-ph/0004014]\,; \,\,\,
E.~Iancu, A.~Leonidov and L.~D.~McLerran,
{\it  Phys.\ Lett.}\,  {\bf B510}, 133 (2001);
[arXiv:hep-ph/0102009];\,\, {\it  Nucl.\ Phys.}\,  {\bf A692}, 583
(2001),
[arXiv:hep-ph/0011241];\,\,\,
E.~Ferreiro, E.~Iancu, A.~Leonidov and L.~McLerran,
 {\it  Nucl.\ Phys.}\  {\bf A703}, 489 (2002),
  [arXiv:hep-ph/0109115];\,\,\,
H.~Weigert,
{\it  Nucl.\ Phys.}  {\bf A703}, 823 (2002),
[arXiv:hep-ph/0004044].
\bibitem{B}
I.~Balitsky,
[arXiv:hep-ph/9509348];\,\,
{\it Phys.\ Rev.} {\bf D60}, 014020 (1999)
[arXiv:hep-ph/9812311]\,\,\,\,
\bibitem{K}
Y.~V.~Kovchegov,
{\it Phys.\ Rev.}  {\bf D60}, 034008  (1999),
[arXiv:hep-ph/9901281].
\bibitem{IIM}
 E.~Iancu, K.~Itakura, L.~McLerran,
  Nucl.\ Phys.\  {\bf A708 } (2002)  327-352.
  [hep-ph/0203137]

\bibitem{BFKL}
 E. A. Kuraev, L. N. Lipatov, and F. S. Fadin, {\it  Sov. Phys.
JETP}
                {\bf 45}, 199 (1977); \,\,\,
Ya. Ya. Balitsky and L. N. Lipatov,
               {\it   Sov. J. Nucl. Phys.}\, {\bf 28}, 22 (1978).

\bibitem{LTAA}
  E.~Levin, K.~Tuchin,
  Nucl.\ Phys.\  {\bf A693 } (2001)  787-798.
  [hep-ph/0101275].
\bibitem{GSEXP}
 A.~M.~Stasto, K.~J.~Golec-Biernat, J.~Kwiecinski,
  Phys.\ Rev.\ Lett.\  {\bf 86 } (2001)  596-599,
  [hep-ph/0007192];\,\,\,L.~McLerran, M.~Praszalowicz,
  Acta Phys.\ Polon.\  {\bf B42 } (2011)  99,
  [arXiv:1011.3403 [hep-ph]]  {\bf B41 } (2010)  1917-1926,
  [arXiv:1006.4293 [hep-ph]].\,\,\, M.~Praszalowicz, [arXiv:1104.1777 [hep-ph]],
  [arXiv:1101.0585 [hep-ph]].
\bibitem{KOWI}
 A.~Kovner and U.~A.~Wiedemann,
  Phys.\ Lett.\  B {\bf 551} (2003) 311
  [arXiv:hep-ph/0207335];
  Phys.\ Rev.\  D {\bf 66} (2002) 034031
  [arXiv:hep-ph/0204277];\,
  Phys.\ Rev.\  D {\bf 66} (2002) 051502
  [arXiv:hep-ph/0112140].
\bibitem{LT}  
~E.~Levin and K.~Tuchin, {\it Nucl.\ Phys.}\,\,{\bf A691}  (2001) 
779,[arXiv:hep-ph/0012167];\,
{\bf B573} (2000) 833,  [arXiv:hep-ph/9908317].
\bibitem{DGLAP}
 V. N. Gribov and L. N. Lipatov, {\it Sov. J. Nucl. Phys} {\bf 15} (1972)
                438;\\
 G. Altarelli and G. Parisi, {\it Nucl. Phys.} {\bf B 126} (1977) 298; \\
Yu. l. Dokshitser, {\it Sov. Phys. JETP} {\bf 46}  (1977) 641.

\bibitem{MUPE}
S.~Munier and R.~B.~Peschanski,
  Phys.\ Rev.\  D {\bf 70} (2004) 077503
  [arXiv:hep-ph/0401215];\,\,
Phys.\ Rev.\  D {\bf 69} (2004) 034008
  [arXiv:hep-ph/0310357];\,\,
  Phys.\ Rev.\ Lett.\  {\bf 91} (2003) 232001
  [arXiv:hep-ph/0309177].
\bibitem{MUTR}
A.~H.~Mueller and D.~N.~Triantafyllopoulos,
{\it Nucl.\ Phys.} \, {\bf B640} (2002) 331
[arXiv:hep-ph/0205167];\,\,D.~N.~Triantafyllopoulos,
{\it Nucl.\ Phys.}\,  {\bf B648} (2003) 293
[arXiv:hep-ph/0209121].
\bibitem{MATH}
Andrei D. Polyanin and Valentin F. Zaitsev,
{\it `` Handbook of nonlinear Partial Differential Equations"},  Chapman $\&$ Hall/CRC, 2004.


\bibitem{KOVLU}
T.~Altinoluk, A.~Kovner and M.~Lublinsky,
  JHEP {\bf 0903} (2009) 110
  [arXiv:0901.2560 [hep-ph]]; JHEP {\bf 0903} (2009) 109
  [arXiv:0901.2559 [hep-ph];\,\,\,
  A.~Kovner and M.~Lublinsky,
  JHEP {\bf 0611} (2006) 083
  [arXiv:hep-ph/0609227];
  Nucl.\ Phys.\  A {\bf 767} (2006) 171
  [arXiv:hep-ph/0510047];
  Phys.\ Rev.\  D {\bf 72} (2005) 074023
  [arXiv:hep-ph/0503155];
  Phys.\ Rev.\ Lett.\  {\bf 94} (2005) 181603
  [arXiv:hep-ph/0502119];
  JHEP {\bf 0503} (2005) 001
  [arXiv:hep-ph/0502071];

\bibitem{BRN}
M.~A.~Braun,
{\it   Phys.\ Lett.}\,  {\bf B632} (2006) 297
  [arXiv:hep-ph/0512057];\,\,
arXiv:hep-ph/0504002\,;
{ \it Eur.\ Phys.\ J.}  {\bf C16}, 337 (2000)
[arXiv:hep-ph/0001268];\,\,\,
  Phys.\ Lett.\ B {\bf 483} (2000) 115
  [arXiv:hep-ph/0003004];\,\,
  Eur.\ Phys.\ J.\ C {\bf 33} (2004) 113
  [arXiv:hep-ph/0309293];\,\,\,
{\it Eur.\ Phys.\ J.}  {\bf C6}, 321 (1999)
[arXiv:hep-ph/9706373];\,\,\,
M.~A.~Braun and G.~P.~Vacca,
{\it Eur.\ Phys.\ J.}  {\bf C6}, 147 (1999)
[arXiv:hep-ph/9711486].

\bibitem{BART}
J.~Bartels, M.~Braun and G.~P.~Vacca,
 { \it Eur.\ Phys.\ J.}  {\bf C40}, 419 (2005)
  [arXiv:hep-ph/0412218]\,;\,\,\,
J.~Bartels and C.~Ewerz,
{\it JHEP} {\bf 9909}, 026 (1999)
[arXiv:hep-ph/9908454]\,;\,\,\,
J.~Bartels and M.~Wusthoff,
{\it Z.\ Phys.} {\bf C66}, 157 (1995)\,;\,\,\,
\,\,\,\,A.~H.~Mueller and B.~Patel,
{\it Nucl.\ Phys.}  {\bf B425}, 471 (1994)
[arXiv:hep-ph/9403256];\,\,\,
J.~Bartels,
Z.\ Phys.\  {\bf C60}, 471 (1993).
\bibitem{LMP}
  E.~Levin, J.~Miller and A.~Prygarin,
  Nucl.\ Phys.\  A {\bf 806} (2008) 245
  [arXiv:0706.2944 [hep-ph]].


\bibitem{LERYA}
E.~Levin  and ~M.~G.~Ryskin,,
{\it Sov. J. Nucl. Phys.} \,{\bf 41}, (1985), 300
 [{\it Yad.\ Fiz.}\,{\bf 41}, (1985)  472].
\bibitem{QIU}
J. ~w. ~Qiu,
 {\it Nucl.\ Phys.}\,{ \bf  B 291},( 1987)  74.
\bibitem{KOLE}
A.~Kormilitzin, E.~Levin,
  Nucl.\ Phys.\  {\bf A849 } (2011)  98-119.
  [arXiv:1009.1468 [hep-ph]].
\bibitem{WS}
  C.~W.~De Jager, H.~De Vries, C.~De Vries,
  Atom.\ Data Nucl.\ Data Tabl.\  {\bf 14 } (1974)  479-508.
\bibitem{WK}
 G.~Watt, H.~Kowalski,
  Phys.\ Rev.\  {\bf D78}, 014016 (2008),
  [arXiv:0712.2670 [hep-ph]];\,\,\,H.~Kowalski, L.~Motyka, G.~Watt,
  Phys.\ Rev.\  {\bf D74 } (2006)  074016,
  [hep-ph/0606272].

\end{thebibliography}
\end{document}